\input phyzzx
\def\T{{\cal T}}
\def\Dslash{D\kern-0.15em\raise0.17ex\llap{/}\kern0.15em\relax}
\def\DslashT{\Dslash^T}
\def\DslashD{\Dslash^\dagger}
\def\DslashDT{{\Dslash^\dagger}^T}
\def\psibar{\overline\psi}
\def\phibar{\overline\phi}
\def\psiT{\psi^T}
\def\psibarT{{\overline\psi}^T}
\def\phiT{\phi^T}
\def\phibarT{{\overline\phi}^T}
\def\CD{C_D}
\def\MD{M^\dagger}
\def\Mprime{M'}
\def\MprimeD{{M'}^\dagger}
\def\Gammae{\Gamma_{11}}
\def\Dslashlr{\mathop{\Dslash}\limits^\leftrightarrow}
\def\DslashTlr{\mathop{\DslashT}\limits^\leftrightarrow}
\def\Dslashl{\mathop{\Dslash}\limits^\leftarrow}
\def\DslashTl{\mathop{\DslashT}\limits^\leftarrow}
\def\partialmul{\mathop{\partial_\mu}\limits^\leftarrow}
\def\Dslashly{\mathop{\Dslash_y}\limits^\leftarrow}
\def\DslashTly{\mathop{\DslashT_y}\limits^\leftarrow}
\def\rslash{\partial\kern-0.026em\raise0.17ex\llap{/}%
          \kern0.026em\relax}
\def\rslashl{\mathop{\rslash}\limits^\leftarrow}
\def\diag{\mathop{\rm diag}}
\def\sqr#1#2{{\vcenter{\hrule height.#2pt
      \hbox{\vrule width.#2pt height#1pt \kern#1pt
          \vrule width.#2pt}
      \hrule height.#2pt}}}

\def\square{{\mathchoice{\sqr84}{\sqr84}{\sqr{5.0}3}{\sqr{3.5}3}}}
\def\kslash{k\kern-0.026em\raise0.17ex\llap{/}%
          \kern0.026em\relax}
\def\qslash{q\kern-0.026em\raise0.17ex\llap{/}%
          \kern0.026em\relax}
\def\Lambdato{{\buildrel{\Lambda\to\infty}\over=}}
\def\Aslash{A\kern-0.026em\raise0.17ex\llap{/}%
          \kern0.026em\relax}
%
\REF\FRO{
S. A. Frolov and A. A. Slavnov,
Phys.\ Lett.\ B309 (1993), 344.}
\REF\FUJ{
K. Fujikawa,
Nucl.\ Phys.\ B428 (1994), 169; Indian J.\ Phys.\ 70A (1996), 275.}
\REF\NAR{
R. Narayanan and H. Neuberger,
Phys.\ Lett.\ B302 (1993), 62.}
\REF\AOK{
S. Aoki and Y. Kikukawa,
Mod.\ Phys.\ Lett.\ A8 (1993), 3517.}
\REF\CHA{
L. N. Chang and C. Soo, Phys.\ Rev.\ D55 (1997), 2410.}
\REF\PAU{
W. Pauli and F. Villars,
Rev.\ Mod.\ Phys.\ 21 (1949), 434.\hfill\break
S. N. Gupta, Proc.\ Phys.\ Soc.\ A66 (1953), 129.}
\REF\WES{
J. Wess and B. Zumino,
Phys.\ Lett.\ 37B (1971), 95.\hfil\break
W. A. Bardeen and B. Zumino,
Nucl.\ Phys.\ B244 (1984), 421.}
\REF\ADL{
S. L. Adler,
Phys.\ Rev.\ 177 (1969), 2426.\hfil\break
J. S. Bell and R. Jackiw,
Nuovo Cim.\ 60A (1969), 47.\hfil\break
W. A. Bardeen,
Phys.\ Rev.\ 184 (1969), 1848.}
\REF\FUJI{
K. Fujikawa,
Phys.\ Rev.\ D29 (1984), 285.}
\REF\THO{
G. 't Hooft,
Phys.\ Rev.\ Lett.\ 37 (1976), 8.}
\REF\CRE{
R. Crewther,
Phys.\ Rev.\ Lett.\ 28 (1972), 1421.\hfil\break
M. Chanowitz and J. Ellis,
Phys.\ Lett.\ 40B (1972), 397; Phys.\ Rev.\ D7 (1973), 2490.}
\REF\FUJIK{
K. Fujikawa,
Phys.\ Rev.\ D25 (1982), 2584;
Phys.\ Rev.\ D21 (1980), 2848; D22 (1980), 1499(E);
Phys.\ Rev.\ Lett.\ 42 (1979), 1195.}
\REF\FUJIKA{
K. Fujikawa,
Phys.\ Rev.\ Lett.\ 44 (1980), 1733; Phys.\ Rev.\ D23 (1981), 2262.}
\REF\NARA{
R. Narayanan and H. Neuberger,
Phys.\ Rev.\ Lett.\ 71 (1993), 3251;
Nucl.\ Phys.\ B412 (1994), 574; Nucl.\ Phys.\ B443 (1995), 305.}
\REF\OURS{
K. Okuyama and H. Suzuki, Phys.\ Lett.\ B382 (1996), 117.}
\REF\ALV{
L. Alvarez-Gaum\'e and P. Ginsparg,
Nucl.\ Phys.\ B243 (1984), 449, and
references therein.}
\REF\SLA{
S. A. Frolov and A. A. Slavnov,
Nucl.\ Phys.\ B411 (1994), 647.\hfil\break
A. A. Slavnov,
Phys.\ Lett.\ B319 (1993), 231; Phys.\ Lett.\ B348 (1995), 553.} 
\REF\NEW{
K. Haga, H. Igarashi, K. Okuyama and H. Suzuki,
Phys.\ Rev.\ D55 (1997), 5325.}
%
\overfullrule=0pt
\pubnum={IU-MSTP/8; hep-th/9603062}
\date={June 1997}
\titlepage
\title{Gauge Invariant Pauli-Villars Regularization of
Chiral Fermions}
\author{%
Kiyoshi Okuyama\foot{%
e-mail: okuyama@mito.ipc.ibaraki.ac.jp}
and Hiroshi Suzuki\foot{%
e-mail: hsuzuki@mito.ipc.ibaraki.ac.jp}}
\address{%
Department of Physics, Ibaraki University, Mito 310, Japan}
\abstract{%
We extend the idea of the generalized Pauli-Villars regularization of
Frolov and Slavnov and analyze the general structure of the
regularization scheme. The gauge anomaly-free condition emerges in a
simple way in the scheme, and, under the standard prescription for
the momentum assignment, the Pauli-Villars Lagrangian provides a
gauge invariant regularization of chiral fermions in arbitrary
anomaly-free representations. The vacuum polarization tensor is
transverse, and the fermion number and the conformal anomalies have
gauge invariant forms. We also point out that the real representation
can be treated in a straightforward manner and the covariant
regularization scheme is directly implemented.}
\endpage
\chapter{Introduction}

In this paper, we extend the idea of the generalized Pauli-Villars
(PV) regularization of chiral fermions proposed by Frolov and Slavnov
some years ago~[\FRO]. We analyze the structure of the regularization
scheme on the basis of a regularization of composite current
operators, as has been performed~[\FUJ] for the generalized PV
regularization proposed by Narayanan and Neuberger~[\NAR]. This type
of analysis provides a simple and transparent way to see the
structure of the regularization of fermion one-loop diagrams.

In the past, several studies regarding this proposal were also
performed: A characterization from the viewpoint of the analytic
index~[\NAR], a verification of the Ward-Takahashi (WT) identities
and an evaluation of the fermion number anomaly by a direct use of
Feynman diagrams~[\AOK] and, more recently the generalization to
arbitrary anomaly-free complex gauge representations in curved
space-time~[\CHA]. We believe our formulation in this paper also
provides a unified view concerning these results.

A regularization based on a diagrammatical calculation, such as the
PV regularization~[\PAU], in general, preserves the Bose symmetry
among external gauge vertices; thus it gives rise to the
consistent~[\WES] gauge anomaly~[\ADL]. Since the consistent anomaly
is not covariant nor invariant under gauge transformation on the
external gauge fields, a Bose symmetric {\it gauge invariant\/}
regularization of chiral fermions, if possible, exists only for
anomaly-free gauge representations. How this anomaly-free requirement
emerges in the scheme is the main concern in the gauge invariant
regularization of chiral fermions. As we will see throughout this
article, the anomaly-free condition emerges in a simple way in the
extension of the generalized PV scheme of Ref.~[\FRO]. This is the
interesting and important property of the proposal.

The organization of this paper is as follows: In~\S2, we extract the
essence of the Lagrangian given in Ref.~[\FRO], which was originally
constructed only for the spinor representation of the SO(10) gauge
group. We then present a general framework for extension to other
gauge representations. In~\S3, the regularized form of the composite
current operators, namely the gauge current, the vector U(1) current
and the axial U(1) current, and the trace part of the energy-momentum
tensor are summarized.

Based on the above setting, real gauge representations are studied in
detail in~\S4. This case allows a straightforward treatment because
of the anomaly-free nature of the representation. We also point out
that the resultant regularized operators are nothing but those in the
covariant regularization scheme in Ref.~[\FUJI].

In~\S5, the complex gauge representation, which is important in view
of applications, is studied. This part of the paper is essentially
the result given in Ref.~[\CHA] specialized to flat space-time, but
we include it for the sake of completeness and for comparison with
the real representation case.

In~\S6, as an illustration of our formulation, the fermion
contribution to the vacuum polarization tensor~[\FRO,\NAR,\AOK,\FUJ]
is calculated for arbitrary anomaly-free representations and for
arbitrary regulator functions. In~\S7, we evaluate the fermion
number anomaly~[\THO] and the conformal anomaly~[\CRE] within our
formulation. We obtain the covariant (or gauge
invariant)~[\FUJIK,\WES] anomalies.

We comment briefly on the relation of the ``Weyl formulation''~[\FRO]
and the vector-like formulation~[\NAR] in~\S8. The final section is
devoted to conclusions.

Throughout this article, we work in Euclidean spacetime,
$ix^0=x^4$, $A_0=iA_4$, $i\gamma^0=\gamma^4$ and
$\gamma_5=i\gamma^0\gamma^1\gamma^2\gamma^3=
\gamma^4\gamma^1\gamma^2\gamma^3$, and in particular,
${\gamma^\mu}^\dagger=-\gamma^\mu$, $\gamma_5^\dagger=\gamma_5$,
$g_{\mu\nu}=-\delta_{\mu\nu}$ and $\varepsilon^{1234}=1$.
\nobreak
\chapter{Generalized PV Lagrangian}

The PV Lagrangian due to Frolov and Slavnov~[\FRO] can be generalized
as follows:
$$
   {\cal L}=\psibar i\Dslash\psi
             -{1\over2}\psibar M\CD\psibarT
             +{1\over2}\psiT\CD^\dagger\MD\psi
   +\phibar Xi\Dslash\phi
             -{1\over2}\phibar\Mprime\CD\phibarT
             +{1\over2}\phiT\CD^\dagger\MprimeD\phi
\eqn\one
$$
In~\one, $\psi$ and~$\phi$ are the fermionic and bosonic Dirac
spinors,\foot{%
We have added free left-handed (spectator) spinors to the Lagrangian
in Ref.~[\FRO]. We will only consider regularization for a
{\it single\/} chiral fermion.}
respectively, each of which possessing the gauge and an internal
space indices. We shall call this internal space ``generation.'' The
first generation component of~$\psi$, $\psi_0$, is the original
massless fermion to be regularized, and other components of $\psi$
and~$\phi$ are massive regulator fields. The number of generations
may be infinite~[\FRO]. $C_D$~is the Dirac charge conjugation
matrix,\foot{%
$\CD\gamma^\mu\CD^{-1}=-{\gamma^\mu}^T$,
$\CD\gamma_5\CD^{-1}=\gamma_5^T$,
$\CD^\dagger=\CD^{-1}$ and $\CD^T=-\CD$.}
and the covariant derivative~$\Dslash$ is defined by
$$
   \Dslash\equiv\gamma^\mu(\partial_\mu-igA_\mu^a\T^aP_R),\quad
   P_R\equiv{1+\gamma_5\over2},
\eqn\two
$$
where $\T^a$~is in general a reducible representation of the gauge
group.

The mass matrices $M$ and~$M'$ in~\one\ possess gauge and the
generation indices. From the statistics of the fields, these matrices
must satisfy
$$
   M^T=M,\quad {M'}^T=-M'.
\eqn\three
$$
The matrix $X$, which also has gauge and the generation indices, has
been introduced to avoid the appearance of tachyonic bosons. Such a
tachyonic state leads to an unwanted pole singularity in the
regulator function~$f(t)$ (see~(4.5)). Since the mass squared of the
bosonic fields is given by $-X^{-1}M'{X^T}^{-1}M'^\dagger$, if
$$
   M'=-XM'X^T,
\eqn\four
$$
then the mass squared is positive definite, $M'M'^\dagger$. The
hermiticity of the action, on the other hand, requires
$$
   X^\dagger=X,\quad[\T^a,X]=0.
\eqn\five
$$
Finally, the mass matrices should satisfy
$$
   \T^aM=-M{\T^a}^T,\quad \T^aM'=-M'{\T^a}^T
\eqn\six
$$
to be gauge invariant. Once a certain set of matrices, $M$, $M'$
and~$X$, which satisfy \three--\six, and a suitable gauge
generator~$\T^a$, which is reduced to the original
representation~$T^a$ on $\psi_0$, are found, the gauge invariant
Lagrangian~\one\ may be constructed. Our general setting~\one\ allows
various extensions of Ref.~[\FRO] which will be discussed in the
subsequent sections.

To reformulate the generalized PV regularization as a regularization
of composite current operators, we need the formal propagators of
$\psi$ and~$\phi$ in a fixed background gauge field. We introduce a
two component notation
$$
   \Psi={\psi\choose\psibarT},\quad
   \Phi={\phi\choose\phibarT}.
\eqn\seven
$$
In terms of these variables, the Lagrangian~\one\ is written as
$$
  {\cal L}=
   {1\over2}\Psi^T\pmatrix{
            \CD^\dagger\MD&-i\DslashT\cr
            i\Dslash&-M\CD\cr}\Psi
   +{1\over2}\Phi^T\pmatrix{
            \CD^\dagger\MprimeD&i\DslashT X^T\cr
            iX\Dslash&-\Mprime\CD\cr}\Phi,
\eqn\eight
$$
where the transpose of the covariant derivative is defined by
$$
   \DslashT\equiv(-\partial_\mu-igA_\mu^a{\T^a}^TP_R^T)
   {\gamma^\mu}^T.
\eqn\nine
$$
This definition (and analogous definitions of $\DslashD$
and~$\DslashDT$ below) is motivated by the matrix notation in the
functional space. Namely,
$$
   \Dslash(x,y)\equiv
    \gamma^\mu(\partial_\mu^x-igA_\mu^a(x){\T^a}P_R)\delta(x-y),
\eqn\ten
$$
and thus
$$
\eqalign{
   \DslashT(x,y)&\equiv
   (\partial_\mu^y-igA_\mu^a(y){\T^a}^TP_R^T)
   {\gamma^\mu}^T\delta(y-x)
\cr
   &=(-\partial_\mu^x-igA_\mu^a(x){\T^a}^TP_R^T)
   {\gamma^\mu}^T\delta(x-y).
\cr
}
\eqn\eleven
$$

Once writing the Lagrangian in the form~\eight, it is straightforward
to find the propagator in a background gauge field~$A_\mu^a$:\foot{%
In deriving these formulas, it is necessary to use relations such as
$\DslashD\CD^\dagger M=\CD^\dagger M\DslashT$,
$\DslashDT\CD^\dagger\MD=\CD^\dagger\MD\Dslash$, etc. These follow
{}from the gauge invariance of the mass term,~\six.}
$$
   \VEV{T\Psi(x)\Psi^T(y)}=
   \left(\matrix{
   -M\CD
   {\displaystyle 1\over\displaystyle\DslashT\DslashDT+\MD M}&
   i\DslashD
   {\displaystyle 1\over\displaystyle\Dslash\DslashD+M\MD}\cr
   -i\DslashDT
   {\displaystyle 1\over\displaystyle\DslashT\DslashDT+\MD M}&
   \CD^\dagger\MD{\displaystyle 1\over\displaystyle\Dslash\DslashD
                  +M\MD}
   \cr
    }\right)\delta(x-y),
\eqn\twelve
$$
and
$$
\eqalign{
   &\VEV{T\Phi(x)\Phi^T(y)}
\cr
   &\qquad=
   \left(\matrix{
   -\Mprime\CD
   {\displaystyle1\over
    \displaystyle\DslashT\DslashDT+\MprimeD\Mprime}&
   i\DslashD X^{-1}
   {\displaystyle1\over
    \displaystyle\Dslash\DslashD+\Mprime\MprimeD}\cr
   i{X^{-1}}^T\DslashDT
   {\displaystyle1\over
    \displaystyle\DslashT\DslashDT+\MprimeD\Mprime}&
   \CD\MprimeD
   {\displaystyle1\over
    \displaystyle\Dslash\DslashD+\Mprime\MprimeD}\cr
    }\right)\delta(x-y).
\cr
}
\eqn\thirteen
$$
In the above expressions, we have introduced
$$
\eqalign{
   \DslashD&\equiv(\partial_\mu-igA_\mu^a\T^aP_R)\gamma^\mu
\cr
   &=\gamma^\mu(\partial_\mu-igA_\mu^a\T^aP_L)
   \ne\Dslash,
\cr
}
\eqn\fourteen
$$
where~$P_L\equiv(1-\gamma_5)/2$, and
$$
   \DslashDT\equiv-{\gamma^\mu}^T
   (\partial_\mu+igA_\mu^a{\T^a}^TP_R^T).
\eqn\fifteen
$$
It is interesting to note that the hermitian conjugate of the
covariant derivative automatically emerges in the inverse operator.
\nobreak
\chapter{Composite current operators in PV regularization}

The central quantity in our analysis is the gauge current, whose
classical form is defined by a functional derivative of the
action~\one\ with respect to the gauge field:
$$
\eqalign{
   &J^{\mu a}(x)
\cr
   &\equiv
   {1\over2}\Psi^T\pmatrix{
            0&-{\T^a}^TP_R^T{\gamma^\mu}^T\cr
            \gamma^\mu\T^aP_R&0\cr}\Psi
   +{1\over2}\Phi^T\pmatrix{
            0&{\T^a}^TP_R^T{\gamma^\mu}^TX^T\cr
            X\gamma^\mu\T^aP_R&0\cr}\Phi.
\cr
}
\eqn\sixteen
$$
Therefore in the PV regularization by Frolov and Slavnov, the
regularized gauge current is defined by
$$
\eqalign{
   &\VEV{J^{\mu a}(x)}_{PV}
\cr
   &\equiv
   {1\over2}\lim_{y\to x}\tr\Biggl[
   (-1)\pmatrix{
            0&-{\T^a}^TP_R^T{\gamma^\mu}^T\cr
            \gamma^\mu\T^aP_R&0\cr}
   \VEV{T\Psi(x)\Psi^T(y)}
\cr
   &\qquad\qquad\qquad\qquad\qquad\qquad
    +\pmatrix{
            0&{\T^a}^TP_R^T{\gamma^\mu}^TX^T\cr
            X\gamma^\mu\T^aP_R&0\cr}
   \VEV{T\Phi(x)\Phi^T(y)}
   \Biggr]
\cr
   &=\lim_{y\to x}\tr\Biggl[
   \gamma^\mu\T^aP_R
   i\DslashD
   {\displaystyle-1\over
    \displaystyle\Dslash\DslashD+M\MD}\delta(x-y)
\cr
   &\qquad\qquad\qquad\qquad\qquad\qquad\qquad\quad
    +\gamma^\mu\T^aP_R
   i\DslashD
  {\displaystyle1\over\displaystyle\Dslash\DslashD+\Mprime\MprimeD}
   \delta(x-y)
   \Biggr],
\cr
}
\eqn\seventeen
$$
where the trace is taken over the generation, gauge and Dirac
indices. We note that this definition is in accord with the standard
Feynman diagrammatical calculation: Further derivatives
of~\seventeen\ with respect to the background gauge field gives a
multi-point one loop vertex function. We will later illustrate such a
calculation of the vacuum polarization tensor. Equation~\seventeen\
therefore summarizes the structure of the regularization scheme in a
neat way.

Strictly speaking, a PV Lagrangian such as~\one\ alone cannot
definitely specify the regularization scheme. As is well-known, one
must supplement the following prescriptions~[\PAU] to the
Lagrangian: 1)~The integrand of the momentum integration must be
summed over all the generations prior to the momentum integration.
2)~The momentum assignment for all the fields (the original fermion
and the regulators) should be taken the same. It is thus important
in~\seventeen\ that the trace over the generation index be taken
before the equal point limit~$y\to x$, according to prescription~1).
Equation~\seventeen\ is as it stands a {\it formal\/} implemention of
prescription~2): The momentum assignment is common for all the
generations. However, one should always be careful in the uniform
momentum assignment in actual calculations such as~(6.4). Those two
underlying prescriptions in the PV regularization are understood
throughout this paper. With this caution in mind, we use the term
``Lagrangian level regularization.''

Another important composite current operator in the chiral gauge
theory is the fermion number current. We define it as the Noether
current associated with a global U(1) rotation~[\AOK,\FUJ]:
$$
\eqalign{
   &\psi(x)\to e^{i\alpha}\psi(x),\quad
    \psibar(x)\to\psibar(x)e^{-i\alpha},
\cr
   &\phi(x)\to e^{i\alpha}\phi(x),\quad
    \phibar(x)\to\phibar(x)e^{-i\alpha},
\cr
}
\eqn\eighteen
$$
or, in terms of the two component notation,
$$
   \Psi(x)\to\pmatrix{e^{i\alpha}&0\cr
                      0&e^{-i\alpha}\cr}\Psi(x),\quad
   \Phi(x)\to\pmatrix{e^{i\alpha}&0\cr
                      0&e^{-i\alpha}\cr}\Phi(x).
\eqn\nineteen
$$
By localizing the infinitesimal parameter~$\alpha$, the Noether
current is defined by
$$
   {\cal L}\to{\cal L}-(\partial_\mu\alpha)J^\mu(x)-\alpha B(x),
\eqn\twenty
$$
where
$$
   J^\mu(x)\equiv
   {1\over2}\Psi^T\pmatrix{
            0&-{\gamma^\mu}^T\cr
            \gamma^\mu&0\cr}\Psi
   +{1\over2}\Phi^T\pmatrix{
            0&{\gamma^\mu}^TX^T\cr
            X\gamma^\mu&0\cr}\Phi,
\eqn\twentyone
$$
and the explicit breaking part~$B(x)$ is given by
$$
    B(x)\equiv
   {1\over2}\Psi^T\pmatrix{
            -2i\CD^\dagger\MD&0\cr
            0&-2iM\CD\cr}\Psi
   +{1\over2}\Phi^T\pmatrix{
            -2i\CD^\dagger\MprimeD&0\cr
            0&-2i\Mprime\CD\cr}\Phi.
\eqn\twentytwo
$$
By defining the composite current operator by the propagators, we
have from~\twentyone,
$$
\eqalign{
   &\VEV{J^\mu(x)}_{PV}
\cr
   &\equiv
   {1\over2}\lim_{y\to x}\tr\Biggl[
   (-1)\pmatrix{
            0&-{\gamma^\mu}^T\cr
            \gamma^\mu&0\cr}
   \VEV{T\Psi(x)\Psi^T(y)}
\cr
   &\qquad\qquad\qquad\qquad\qquad\qquad\qquad\quad
   +\pmatrix{
            0&{\gamma^\mu}^TX^T\cr
            X\gamma^\mu&0\cr}
   \VEV{T\Phi(x)\Phi^T(y)}
   \Biggr]
\cr
   &=\lim_{y\to x}\tr\Biggl[
   \gamma^\mu
   i\DslashD
   {\displaystyle-1\over
    \displaystyle\Dslash\DslashD+M\MD}\delta(x-y)
\cr
   &\qquad\qquad\qquad\qquad\qquad\qquad\qquad\quad
    +\gamma^\mu
   i\DslashD
  {\displaystyle1\over\displaystyle\Dslash\DslashD+\Mprime\MprimeD}
   \delta(x-y)
   \Biggr].
\cr
}
\eqn\twentythree
$$
If the composite current operator is well regularized, we may derive
the corresponding WT identity,
$$
   \partial_\mu\VEV{J^\mu(x)}_{PV}=\VEV{B(x)}_{PV},
\eqn\twentyfour
$$
as a result of the naive equation of motion. Under a Lagrangian
level regularization, a possible anomaly associated with a certain
global symmetry should arise as an {\it explicit\/} symmetry breaking
term in the Lagrangian. Thus the vacuum expectation value of~$B(x)$,
the right-hand side of~\twentyfour, gives rise to the fermion number
anomaly. We will later verify that this is, in fact, the case. This
situation is analogous to an evaluation of the gauge anomaly in
chiral gauge theories by the {\it conventional\/} PV regularization,
where the {\it gauge\/} symmetry is explicitly broken by the PV mass
term.

We may equally well use another definition of the regularized fermion
number current. It is defined by the Noether current associated with
a global {\it axial\/} U(1) rotation~[\AOK,\FUJ]:
$$
\eqalign{
   &\psi(x)\to e^{i\alpha\gamma_5}\psi(x),\quad
    \psibar(x)\to\psibar(x)e^{i\alpha\gamma_5},
\cr
   &\phi(x)\to e^{i\alpha\gamma_5}\phi(x),\quad
    \phibar(x)\to\phibar(x)e^{i\alpha\gamma_5}.
\cr
}
\eqn\twentyfive
$$
By the same procedure as above, we find the associated Noether
current
$$
\eqalign{
   &\VEV{J_5^\mu(x)}_{PV}
\cr
   &=\lim_{y\to x}\tr\Biggl[
   \gamma^\mu\gamma_5
   i\DslashD
   {\displaystyle-1\over
    \displaystyle\Dslash\DslashD+M\MD}\delta(x-y)
    +\gamma^\mu\gamma_5
   i\DslashD
  {\displaystyle1\over\displaystyle\Dslash\DslashD+\Mprime\MprimeD}
   \delta(x-y)
   \Biggr].
\cr
}
\eqn\twentysix
$$
In what follows, we find that it is always possible to choose $M$
and~$M'$, such that composite operators in \twentythree\
and~\twentysix\ are regularized, if the gauge representation is free
of the gauge anomaly. In this case, we can see that the currents
\twentythree\ and~\twentysix\ are the same object: We first note
from \two\ and~\fourteen\ that
$$
   {\displaystyle1\over
    \displaystyle\Dslash\DslashD+M\MD}=
   P_L{\displaystyle1\over
    \displaystyle\Dslash\DslashD+M\MD}+
   P_R{\displaystyle1\over
    \displaystyle\rslash^2+M\MD},
\eqn\twentyseven
$$
holds, as does an analogous relation for the bosonic part. Putting
these into \twentythree\ and~\twentysix\ and noting that there exists
no constant vector independent of~$A_\mu$, we see that only the first
term of~\twentyseven\ survives; consequently \twentythree\
and~\twentysix\ are the same operator (note~$\gamma_5P_R=P_R$). Of
course this is an expected result because only the right-handed
fields are coupled to the background gauge field. After observing the
equivalence of \twentythree\ and~\twentysix, we use~\twentythree\ as
the fermion number current in what follows.

Another interesting operator is the trace part of the
energy-momentum tensor~$T_\mu^\mu(x)$, which is defined in the
original theory by\foot{%
This definition requires some explanation: If one simply uses the
standard definition of the energy-momentum tensor of the spinor
field, $-3$~times our result will be obtained. Our definition,
following Ref.~[\FUJIKA], is specified by the {\it general
coordinate\/} invariance in the background gravitational field
(see Refs.~[\FUJIKA,\FUJ] for more details).}
$$
   \overline\psi_0i\Dslash\psi_0\to
   \overline\psi_0i\Dslash\psi_0-\alpha(x)T_\mu^\mu(x),\quad
   T_\mu^\mu(x)\equiv\overline\psi_0{i\over2}\Dslashlr\psi_0,
\eqn\twentyeight
$$
where the variation of the field is
$$
   \psi_0(x)\to e^{-\alpha(x)/2}\psi_0(x),\quad
   \bar\psi_0(x)\to \bar\psi_0(x)e^{-\alpha(x)/2}.
\eqn\twentynine
$$
By generalizing the rescaling of the field~\twentynine\ to all the
regulator fields, the regularized version of the trace part of the
energy-momentum tensor is defined by
$$
\eqalign{
   &\VEV{T_\mu^\mu(x)}_{PV}
\cr
   &\equiv{1\over2}\lim_{y\to x}\tr\Biggl[
    (-1)\pmatrix{
            0&-{i\over2}\DslashT_x\cr
            {i\over2}\Dslash_x&0\cr}\VEV{T\Psi(x)\Psi^T(y)}
\cr
   &\qquad\qquad\qquad\qquad\qquad\qquad
     +(-1)\VEV{T\Psi(x)\Psi^T(y)}\pmatrix{
            0&{i\over2}\DslashTly\cr
            -{i\over2}\Dslashly&0\cr}
\cr
   &\qquad\qquad\qquad
       +\pmatrix{
            0&{i\over2}\DslashT_xX^T\cr
            {i\over2}X\Dslash_x&0\cr}\VEV{T\Phi(x)\Phi^T(y)}
\cr
   &\qquad\qquad\qquad\qquad\qquad\qquad
    +\VEV{T\Phi(x)\Phi^T(y)}\pmatrix{
            0&-{i\over2}\DslashTly X^T\cr
            -{i\over2}X\Dslashly&0\cr}\Biggr].
\cr
}
\eqn\thirty
$$
In \twentyeight\ and~\thirty, $\Dslashlr=\Dslash-\Dslashl$,
$\DslashTlr=\DslashT-\DslashTl$
and
$$
   \Dslashl\equiv
   \gamma^\mu(\partialmul+igA_\mu^a\T^aP_R),\quad
   \DslashTl\equiv
   (-\partialmul+igA_\mu^a{\T^a}^TP_R^T){\gamma^\mu}^T.
\eqn\thirtyone
$$
Noting $\delta(x-y)\Dslashly=-\Dslash_x\delta(x-y)$ and
$\Dslash^{-1}M\MD=M\MD\Dslash^{-1}$, and thus
$$
   {1\over\Dslash\DslashD+M\MD}\Dslash
   =\Dslash{1\over\DslashD\Dslash+M\MD},
\eqn\thirtytwo
$$
we finally have
$$
\eqalign{
   &\VEV{T_\mu^\mu(x)}_{PV}
\cr
   &={1\over2}\lim_{y\to x}\tr\left[
     \Dslash\DslashD{1\over\Dslash\DslashD+M\MD}\delta(x-y)
     +\Dslash\DslashD{-1\over\Dslash\DslashD+\Mprime\MprimeD}
     \delta(x-y)
     \right]
\cr
   &\qquad+{1\over2}\lim_{y\to x}\tr\left[
     \DslashD\Dslash{1\over\DslashD\Dslash+M\MD}\delta(x-y)
     +\DslashD\Dslash{-1\over\DslashD\Dslash+\Mprime\MprimeD}
     \delta(x-y)
     \right].
\cr
}
\eqn\thirtythree
$$

The composite operators, the gauge current~\seventeen, the fermion
number current~\twentythree, and the trace part of the
energy-momentum tensor~\thirtythree\ will be analyzed in detail in
the following discussion.
\nobreak
\chapter{Real representations}

As was noted in the Introduction, the gauge invariant regularization
of a chiral fermion is possible only for anomaly-free gauge
representations. The situation is simple for real representations,
because the anomaly-free condition is always fulfilled by the
presence of a matrix~$U$ which transforms the original representation
to the adjoint representation. As we will see below, the generalized
PV Lagrangian~\one\ can utilize this fact, and this is an advantage
of the present framework.

For any real representation~$T^a$, there exists a unitary matrix~$U$
such that
$$
   -{T^a}^T=-{T^a}^*=UT^aU^\dagger.
\eqn\thirtyfour
$$
For a real-positive representation, $U$~is symmetric, and for a
pseudo-real representation, $U$~is anti-symmetric. For both cases, we
can make the choice:
$$
   \T^a\equiv T^a\otimes1,\quad
   M=U^\dagger\otimes m,\quad
   M'=U^\dagger\otimes m',\quad
   X=1\otimes x,
\eqn\thirtyfive
$$
where the first index acts on the gauge and the second acts on the
generation index. It turns out that the nature of $m$ and~$m'$ are
quite different depending on whether the representation is
real-positive or pseudo-real. We thus treat them separately.

When the chiral fermion belongs to a real-positive representation of
the gauge group, we can take, for example,
$$
   m=\pmatrix{0&0\cr
              0&\sqrt{2}\cr}\Lambda,\quad
   m'=\pmatrix{0&1\cr
               -1&0\cr}\Lambda,\quad
   x=\pmatrix{1&0\cr
              0&-1\cr},
\eqn\thirtysix
$$
where $\Lambda$~is the cutoff parameter. It is readily verified
that these matrices satisfy \three--\six. Since
$M\MD=\bigl({0\atop0}\,{0\atop2}\bigr)\Lambda^2$ and
$M'\MprimeD=\bigl({1\atop0}\,{0\atop1}\bigr)\Lambda^2$, the
regularized gauge current~\seventeen\ is given by
$$
   \VEV{J^{\mu a}(x)}_{PV}
   =\lim_{y\to x}\tr\left[
            \gamma^\mu T^aP_R\DslashD
            {\displaystyle1\over\displaystyle i\Dslash\DslashD}
      f(\Dslash\DslashD/\Lambda^2)\delta(x-y)\right],
\eqn\thirtyseven
$$
where we have defined the regulator function
$$
   f(t)\equiv{2\over(t+1)(t+2)},
\eqn\thirtyeight
$$
which vanishes rapidly as~$t\to\infty$ and satisfies
$$
\eqalign{
   &f(0)=1,\quad
   \lim_{t\to0}tf'(t)
   =\lim_{t\to0}t^2f''(t)=\lim_{t\to0}t^3f^{(3)}(t)=0,
\cr
   &\lim_{t\to\infty}tf(t)=\lim_{t\to\infty}t^2f'(t)=
    \lim_{t\to\infty}t^2f''(t)=\lim_{t\to\infty}t^3f^{(3)}(t)=0.
\cr
}
\eqn\thirtynine
$$
Due to the rapid damping property in the second line, the gauge
current~\thirtyseven\ is well regularized, and the limit~$y\to x$ can
safely be taken.

Let us next consider the pseudo-real case, for which the matrix~$U$
is anti-symmetric. In~\thirtyfive\ we may choose
$$
   m=\pmatrix{0&  & &  & & \cr
               & 0&2&  & & \cr
               &-2&0&  & & \cr
               &  & &0&4& \cr
               &  & &-4&0& \cr
               &  & &  & &\ddots\cr}\Lambda,\quad
   m'=\pmatrix{0&1& & & \cr
               1&0& & & \cr
                & &0&3& \cr
                & &3&0& \cr
                & & & &\ddots\cr}\Lambda,
\eqn\forty
$$
and
$$
   x=\diag(1,-1,1,-1,\cdots).
\eqn\fortyone
$$
These matrices again satisfy \three--\six. Note that in this case we
have introduced an infinite number of regulator fields. The mass
squared is given by
$$
   M\MD=1\otimes\pmatrix{0^2&   &   & \cr
                             &2^2&   & \cr
                             &   &2^2& \cr
                             &   &   &\ddots\cr}\Lambda^2,\quad
   M'\MprimeD=1\otimes\pmatrix{1^2&   &   & &\cr
                                   &1^2&   & &\cr
                                   &   &3^2& &\cr
                                   &   &   &3^2&\cr
                                   &   &   & &\ddots\cr}\Lambda^2.
\eqn\fortytwo
$$
With the above choice of mass matrices, the gauge current
operator~\seventeen\ becomes
$$
\eqalign{
   \VEV{J^{\mu a}(x)}_{PV}
   &=\lim_{y\to x}\tr\left[
            \gamma^\mu T^aP_R\DslashD
            {\displaystyle1\over\displaystyle i\Dslash\DslashD}
   \sum_{n=-\infty}^\infty
   {\displaystyle(-1)^n\Dslash\DslashD\over
    \displaystyle\Dslash\DslashD+n^2\Lambda^2}\delta(x-y)\right]
\cr
   &=\lim_{y\to x}\tr\left[
            \gamma^\mu T^aP_R\DslashD
            {\displaystyle1\over\displaystyle i\Dslash\DslashD}
      f(\Dslash\DslashD/\Lambda^2)\delta(x-y)\right],
\cr
}
\eqn\fortythree
$$
where the regulator function~$f(t)$ is defined by~[\FRO]
$$
   f(t)\equiv\sum_{n=-\infty}^\infty{(-1)^nt\over t+n^2}=
   {\pi\sqrt{t}\over\sinh(\pi\sqrt{t})},
\eqn\fortyfour
$$
which again has the desired properties~\thirtynine.\foot{%
Although it is not necessarily required, if one prefers the regulator
function~\fortyfour\ in the real-positive case,
$$
   m=\pmatrix{0& & & & & \cr
                               &2& & & & \cr
                               & &2& & & \cr
                               & & &4& & \cr
                               & & & &4& \cr
                               & & & & &\ddots\cr}\Lambda,\quad
   m'=\pmatrix{0 &1&  & & \cr
                               -1&0&  & & \cr
                                 & &0 &3& \cr
                                 & &-3&0& \cr
                                 & &  & &\ddots\cr}\Lambda,
$$
and $x$ in~\fortyone\ may be chosen.}

In \thirtyseven\ and~\fortythree, we see that all the divergences
are made finite gauge invariantly. In fact these expressions are
nothing but those of the covariant regularization scheme~[\FUJI],
which is formulated as a gauge invariant damping factor
$f(\Dslash\DslashD/\Lambda^2)$ insertion in the original fermion
propagator. The covariant regularization is known to give the
covariant anomaly~[\FUJIK,\WES]:\foot{%
The vacuum overlap approach~[\NARA] in the lattice chiral gauge
theory, which is closely related to the generalized PV
regularization~[\NAR], is known to give the consistent anomaly.}
$$
   D_\mu\VEV{J^{\mu a}(x)}_{PV}\Lambdato{ig^2\over32\pi^2}
   \varepsilon^{\mu\nu\rho\sigma}\tr
   \bigl(T^aF_{\mu\nu}F_{\rho\sigma}\bigr),
\eqn\fortyfive
$$
where the field strength is defined by
$F_{\mu\nu}\equiv\left(\partial_\mu A_\nu^a
                       -\partial_\nu A_\mu^a
                       +gf^{abc}A_\mu^bA_\nu^c\right)T^a$
and the right-hand side vanishes due to the anomaly-free condition
$\tr(T^a\{T^b,T^c\})=0$.\foot{%
Hence it trivially satisfies the Wess-Zumino consistency condition.
This is consistent with the fact that we are treating a Lagrangian
level regularization which respects the Bose symmetry among gauge
vertices.}
Starting from \thirtyseven\ or~\fortythree, we can directly evaluate
the gauge anomaly~\fortyfive\ and the calculation is almost identical
to the passage from (7.1) to~(7.6)~[\OURS]. The regularization scheme
due to Frolov and Slavnov therefore gives a Lagrangian level
implementation of the covariant regularization scheme in
Ref.~[\FUJI]. This aspect of the generalized PV regularization is
studied in detail in Ref.~[\OURS].

Going back to consideration of the pseudo-real case, we have chosen
the mass matrices~\forty\ simply because the explicit summation over
the generation index can be performed as~\fortyfour. We may make
other choice of $m$ and~$m'$ in~\thirtyfive, and it may be thought
that a more clever choice could reduce the number of regulator fields
to finite value. We show below, however, that an infinite number of
regulator fields are always needed, at least when relying on the
construction~\thirtyfive.\foot{%
If all the fields belong to the same irreducible representation,
\five\ and Schur's lemma imply the structure~$X=1\otimes x$.}

Let us first assume the number of the generation is finite, and $m$,
$m'$ and~$x$ are finite dimensional matrices. From~\three, $m^T=-m$
(note $U$~is anti-symmetric). Therefore if $m$~is an even dimensional
matrix, it may be put into a block diagonal form by an orthogonal
transformation of~$\psi$:
$$
   m=\pmatrix{m_1\otimes\varepsilon&&&\cr
              &m_2\otimes\varepsilon&&\cr
              &&\ddots&\cr
              &&&m_k\otimes\varepsilon\cr},\quad
   \varepsilon\equiv\pmatrix{0&1\cr-1&0\cr}
\eqn\fortysix
$$
and
$$
   MM^\dagger=1\otimes\pmatrix{m_1^2\otimes I&&&\cr
              &m_2^2\otimes I&&\cr
              &&\ddots&\cr
              &&&m_k^2\otimes I\cr},\quad
   I\equiv\pmatrix{1&0\cr0&1\cr}.
\eqn\fortyseven
$$

The number of massless fermion fields, if such fields exist, is
always even. Since we are constructing a regularization for a
{\it single\/} massless fermion, the matrix~$m$ should be odd
dimensional. On the other hand, from~\five, $x$ is hermitian, and by
a unitary transformation and a rescaling of~$\phi$, it may be put
into the form
$$
   x=\diag\bigl(\overbrace{1,1,\cdots,1}^k,
           \overbrace{-1,-1,\cdots,-1}^l\bigr).
\eqn\fortyeight
$$
Then \four\~implies $m'_{ij}=-x_ix_jm'_{ij}$, and $m'$~has the
structure
$$
   m'=\pmatrix{0&Y\cr Y^T&0\cr}\Lambda,
\eqn\fortynine
$$
where $Y$~is a $k\times l$~matrix. If $k<l$, $\dim\ker Y\geq l-k$,
and if $k>l$, $\dim\ker Y^T\geq k-l$. For both of these cases, there
exists at least one linear combination of~$\phi_i$ which remains
massless. This is an unwanted massless bosonic field, and we should
take~$k=l$. Therefore $x$ and~$m'$ must be even dimensional.

We now have an odd number of fermions and an even number of bosons
in the same gauge representation. However, the PV condition for the
ultraviolet divergence reduction~[\PAU] always requires the same
numbers of fermionic and bosonic degrees of freedom. From the above
argument, this is impossible when the number of the generation is
finite. This shows that, at least within the
construction~\thirtyfive, an infinite number of regulator fields are
always needed. As we have observed, they in fact regularize the
theory.

By the same procedure as for the gauge current, the regularized U(1)
global current~\twentythree\ becomes
$$
   \VEV{J^\mu(x)}_{PV}=\lim_{y\to x}\tr\left[
            \gamma^\mu\DslashD
            {\displaystyle1\over\displaystyle i\Dslash\DslashD}
      f(\Dslash\DslashD/\Lambda^2)\delta(x-y)\right],
\eqn\fifty
$$
with an appropriate regulator function~$f(t)$. Similarly, the trace
part of the energy-momentum tensor~\thirtythree\ is given by
$$
   \VEV{T_\mu^\mu(x)}_{PV}
   ={1\over2}\lim_{y\to x}\tr\left[
     f(\Dslash\DslashD/\Lambda^2)\delta(x-y)
     +f(\DslashD\Dslash/\Lambda^2)\delta(x-y)
     \right].
\eqn\fiftyone
$$

Therefore, for real representations, the generalized PV Lagrangian
provides a complete gauge invariant regularization of the gauge
current \thirtyseven\ or~\fortythree, as well as the fermion number
current~\fifty\ and the trace part of the energy-momentum
tensor~\fiftyone.
\nobreak
\chapter{Complex representations}

The generalized PV regularization~[\FRO] was originally formulated
for the irreducible spinor representation of~SO(10), i.e., an
anomaly-free complex representation, which is important for an
application to the standard model. Quite recently~[\CHA] the
construction has been successfully generalized for arbitrary
anomaly-free complex representations. We include these results in
this section (specializing them to flat space-time) and compare the
situation with that of the real representation in the previous
section.

For a generalization of the PV Lagrangian to arbitrary complex
representations, it is crucial to introduce a doubled
representation~[\CHA]:
$$
   \T^a=\pmatrix{T^a&0\cr0&-T^{a*}\cr}\otimes1.
\eqn\fiftytwo
$$
With this doubling of the gauge representation, the following choice
of matrices in~\one\ satisfies \three--\six:
$$
   M=\sigma^1\otimes m,\quad M'=i\sigma^2\otimes m',\quad
   X=\sigma^3\otimes1,
\eqn\fiftythree
$$
where $\sigma^i$~is the Pauli matrix. The relation to the original
SO(10)~model~[\FRO] is\foot{%
The gamma matrix for the spinor representation
satisfies the Clifford algebra
$\{\Gamma_i,\Gamma_j\}=2\delta_{ij}$ and is hermitian.
The gauge generator is defined by
$\T^a=i[\Gamma_i,\Gamma_j]/2$ and the irreducible representation is
projected by $(1+\Gammae)/2$. The ``chiral'' matrix $\Gammae$ is
defined by $\Gammae=-i\Gamma_1\Gamma_2\cdots\Gamma_{10}$ and
satisfies
$\{\Gammae,\Gamma_i\}=0$, $\Gammae^\dagger=\Gammae$ and
$\Gammae^2=1$. The ``charge conjugation'' matrix $C$ has the
properties $C\Gamma_iC^{-1}=-\Gamma_i^T$,
$C\Gammae C^{-1}=-\Gammae^T$, $C^\dagger=C^{-1}$ and $C^T=-C$.}
$$
   \sigma^1\to\Gamma_{11}C,\quad
   i\sigma^2\to C,\quad\sigma^3\to\Gamma_{11}.
\eqn\fiftyfour
$$
The matrices $m$ and~$m'$ are chosen as
$$
   m=\pmatrix{0& & & \cr
               &2& & \cr
               & &4& \cr
               & & &\ddots\cr}\Lambda,\quad
   m'=\pmatrix{1& & & \cr
                &3& & \cr
                & &5& \cr
                & & &\ddots\cr}\Lambda.
\eqn\fiftyfive
$$
Although the regulator fields must belong to the doubled
representation~\fiftytwo\ to have a non-vanishing mass, the original
massless fermion~$\psi_0$ must be projected by $(1+\sigma^3)/2$ to
have the original complex representation~$T^a$, rather than the
doubled representation~$\T^a$.

Let us now consider the regularized composite operators. For example,
the regularized gauge current~\seventeen\ is given by
$$
\eqalign{
   &\VEV{J^{\mu a}(x)}_{PV}
\cr
   &={1\over2}\lim_{y\to x}\tr\left[
            \gamma^\mu\T^aP_R\DslashD
            {\displaystyle1\over\displaystyle i\Dslash\DslashD}
   \sum_{n=-\infty}^\infty
   {\displaystyle(-1)^n\Dslash\DslashD\over
    \displaystyle\Dslash\DslashD+n^2\Lambda^2}\delta(x-y)\right]
\cr
   &\qquad\qquad\qquad\qquad\qquad\qquad\qquad
    +{1\over2}\lim_{y\to x}\tr\left[
            \gamma^\mu\sigma^3\T^aP_R\DslashD
            {\displaystyle1\over\displaystyle i\Dslash\DslashD}
          \delta(x-y)\right]
\cr
   &={1\over2}\lim_{y\to x}\tr\left[
            \gamma^\mu\T^aP_R\DslashD
            {\displaystyle1\over\displaystyle i\Dslash\DslashD}
      f(\Dslash\DslashD/\Lambda^2)\delta(x-y)\right]
\cr
   &\qquad\qquad\qquad\qquad\qquad\qquad\qquad\quad
     +{1\over2}\lim_{y\to x}\tr\left[
            \gamma^\mu\sigma^3\T^aP_R
            {\displaystyle1\over\displaystyle i\Dslash}
          \delta(x-y)\right],
\cr
}
\eqn\fiftysix
$$
where the regulator function is given by~\fortyfour\ and we have used
the fact that the first generation fermion is projected
by~$(1+\sigma^3)/2$.

Similarly, the fermion number current~\twentythree\ and the trace
part of the energy-momentum tensor~\thirtythree\ become
$$
\eqalign{
   &\VEV{J^\mu(x)}_{PV}
\cr
   &={1\over2}\lim_{y\to x}\tr\left[
            \gamma^\mu\DslashD
            {\displaystyle1\over\displaystyle i\Dslash\DslashD}
      f(\Dslash\DslashD/\Lambda^2)\delta(x-y)\right]
     +{1\over2}\lim_{y\to x}\tr\left[
            \gamma^\mu\sigma^3
            {\displaystyle1\over\displaystyle i\Dslash}
          \delta(x-y)\right]
\cr
}
\eqn\fiftyseven
$$
and
$$
   \VEV{T_\mu^\mu(x)}_{PV}
   ={1\over2}\cdot{1\over2}\lim_{y\to x}\tr\left[
     f(\Dslash\DslashD/\Lambda^2)\delta(x-y)
     +f(\DslashD\Dslash/\Lambda^2)\delta(x-y)
     \right].
\eqn\fiftyeight
$$

In deriving \fiftysix--\fiftyeight, we have taken the
trace over the generation index. In the final line of~\fiftysix\
and of~\fiftyseven, the first term is completely regularized. Since
the regulator function~$f(t)$ damps rapidly enough as~$t\to\infty$,
the limit~$y\to x$ can safely be taken to give the finite result. It
should also be noted that an infinite number of regulator fields are
always needed to balance the fermionic and the bosonic degrees of
freedom~[\FRO]. On the other hand, the last terms of \fiftysix\
and~\fiftyseven\ do not yet have a regulator function, and the
expression is ill-defined in general. Compare these with \fortythree\
and~\fifty\ for the real representations. The origin of the
complication in the complex representation is that it is in general
not anomaly-free, and it is impossible to distinguish the
anomaly-free representations from the anomalous ones at the level of
Lagrangian construction. Therefore we must supplement the
regularization scheme with the anomaly-free condition.

In fact, as first pointed out in Ref.~[\FRO] (see also
Refs.~[\NAR,\AOK]) and generalized in Ref.~[\CHA], these
unregularized terms become finite if and only if $\tr T^a=0$ and
$\tr T^a\{T^b,T^c\}=0$, i.e., if free of the gauge (and Lorentz)
anomaly. To see this, we note the perturbative expansion
$$
   {1\over i\Dslash}
   ={1\over i\rslash}
    +{1\over i\rslash}(-g\Aslash P_R){1\over i\rslash}
    +{1\over i\rslash}(-g\Aslash P_R){1\over i\rslash}(-g\Aslash P_R)
    {1\over i\rslash}
    +\cdots.
\eqn\fiftynine
$$
Using $\delta(x-y)=\int d^4k\,e^{-ik(x-y)}/(2\pi)^4$ and the
momentum representation of~$A_\mu^a(x)$, we see that the last terms
in \fiftysix\ and~\fiftyseven\ generate Feynman integrals with an
insertion of~$\sigma^3$. The momentum integration of the one, two,
three and four-point functions is power-counting divergent, and
higher point functions are convergent. Therefore if
$$
   \tr(\sigma^3\T^{a_1}\cdots\T^{a_n})
  =\tr(T^{a_1}\cdots T^{a_n})+(-1)^{n+1}\tr(T^{a_n}\cdots T^{a_1})=0,
   \quad{\rm for}\quad n\leq4,
\eqn\sixty
$$
the integrand of the power-counting divergent expression
vanishes. As is easily verified~[\AOK,\CHA], this condition is
equivalent to $\tr T^a=0$ and $\tr(T^a\{T^b,T^c\})=0$. In the
framework of Ref.~[\FRO], the anomaly-free complex representation is
distinguished from anomalous ones in this way, and the generalized PV
regularization works only for the anomaly-free case, as should be
the case.\foot{%
When the construction of this section is applied to {\it real\/}
representations~[\CHA], Eq.~\sixty\ holds for {\it all\/}~$n$ due to
the presence of the matrix~$U$.}
\nobreak
\chapter{Vacuum polarization tensor}

As an illustration, we compute the fermion contribution to the vacuum
polarization tensor~[\FRO,\AOK,\FUJ] in our formulation, \fortythree\
or~\fiftysix. For simplicity, we first consider the case of the real
representation~\fortythree\ and comment later on the anomaly-free
complex representation~\fiftysix.

The vacuum polarization tensor is defined by the first functional
derivative of the gauge current~\fortythree\ with respect to the
background gauge field:
$$
   \left.{\delta\VEV{J^{\mu a}(x)}_{PV}
          \over\delta(gA_\nu^b(z))}\right|_{A=0}
   \equiv\int{d^4q\over(2\pi)^4}\,e^{-iq(x-z)}\Pi^{\mu\nu ab}(q).
\eqn\sixtyone
$$
We assume the following form of the regulator function:
$$
   \widetilde f(t)=\sum_n{c_nt\over t+m_n^2/\Lambda^2}.
\eqn\sixtytwo
$$
The above examples, $f(t)$ in~\thirtyeight, $f(t)$ in~\fortyfour, and
the generalized PV in general, are certainly contained in this class
of functions. We also assume that $c_n$ and~$m_n$ are chosen so as to
satisfy~\thirtynine. From the definition of the covariant derivative,
\two\ and~\fourteen, we have
$$
\eqalign{
   &\left.{\delta\VEV{J^{\mu a}(x)}_{PV}
           \over\delta(gA_\nu^b(z))}\right|_{A=0}
\cr
   &=-\tr(T^aT^b)
    \lim_{y\to x}\tr\biggl(P_L\gamma^\mu\biggl\{
    \delta(x-z)\gamma^\nu\sum_n{c_n\over\square+m_n^2}
\cr
   &\qquad\qquad
   +i\rslash\sum_n{c_n\over\square+m_n^2}\left[
   i\rslash\delta(x-z)\gamma^\nu
    +\delta(x-z)\gamma^\nu i\rslash
   \right]{1\over\square+m_n^2}\biggr\}\delta(x-y)\biggr).
\cr
}
\eqn\sixtythree
$$
All the derivatives in~\sixtythree\ act on everything to their right.
We then use the momentum representation of the delta
functions, $\delta(x-z)=\int d^4q\,e^{-iq(x-z)}/(2\pi)^4$~etc. The
vacuum polarization tensor~\sixtyone\ is then given by
$$
\eqalign{
   &\Pi^{\mu\nu ab}(q)
\cr
   &=-\tr(T^aT^b)\int{d^4k\over(2\pi)^4}
   \biggl\{
   \tr(P_L\gamma^\mu\gamma^\nu)\sum_n
   {c_n\over-(k+q)^2+m_n^2}{m_n^2\over-k^2+m_n^2}
\cr
   &\qquad\qquad\qquad\qquad\qquad\quad
   +\tr[P_L\gamma^\mu(\kslash+\qslash)\gamma^\nu\kslash]\sum_n
   {c_n\over-(k+q)^2+m_n^2}{1\over-k^2+m_n^2}\biggr\}.
\cr
}
\eqn\sixtyfour
$$
The subsequent steps are standard: We introduce the Feynman parameter
to combine the denominators, shift the integration momentum,\foot{%
It is important to shift all the integration momenta in the same way.
Otherwise, one obtains a gauge non-invariant (but finite) piece which
is proportional to~$g^{\mu\nu}$.}
and take the trace of the gamma matrices. Noting the definition
of~$\widetilde f(t)$~\sixtytwo, we have
$$
\eqalign{
   &\Pi^{\mu\nu ab}(q)
   =-\tr(T^aT^b)\int_0^1dx\int{d^4k\over(2\pi)^4}
\cr
   &\quad\times\biggl\{
   g^{\mu\nu}
   \biggl[2\sum_n{c_n\over-k^2-q^2x(1-x)+m_n^2}
         +\sum_n{c_nk^2\over[-k^2-q^2x(1-x)+m_n^2]^2}\biggr]
\cr
   &\qquad\qquad\qquad\qquad
   -4x(1-x)(q^\mu q^\nu-g^{\mu\nu}q^2)\sum_n
   {c_n\over[-k^2-q^2x(1-x)+m_n^2]^2}
   \biggr\}
\cr
   &=-{1\over16\pi^2}\tr(T^aT^b)\int_0^1dx\int_0^\infty dt
\cr
   &\quad\times\left[
   \Lambda^2g^{\mu\nu}\bigl(2t+t^2{d\over dt}\bigr)
   {\widetilde f(t+s)\over t+s}
   +4x(1-x)(q^\mu q^\nu-g^{\mu\nu}q^2)t{d\over dt}
   {\widetilde f(t+s)\over t+s}
   \right],
\cr
}
\eqn\sixtyfive
$$
where $t\equiv-k^2/\Lambda^2$ and $s\equiv-q^2x(1-x)/\Lambda^2$.

Now if $\lim_{t\to\infty}t\widetilde f(t)=0$ as in~\thirtynine, the
quadratically divergent gauge non-invariant term disappears after a
partial integration of the first term. Therefore we obtain the final
result
$$
\eqalign{
   &\Pi^{\mu\nu ab}(q)
\cr
   &={1\over4\pi^2}\tr(T^aT^b)
   (q^\mu q^\nu-g^{\mu\nu}q^2)\int_0^1dx\,x(1-x)
   \biggl[
   \int_s^\infty dt\,{f(t)\over t}
   -\int_s^\infty dt\,{f(t)-\widetilde f(t)\over t}   
   \biggr]
\cr
   &\Lambdato
   {1\over24\pi^2}\tr(T^aT^b)(q^\mu q^\nu-g^{\mu\nu}q^2)
   \biggl[\log{\Lambda^2\over-q^2}+{5\over3}-2\log{\pi\over2}
   +\int_0^\infty dt\,{\widetilde f(t)-f(t)\over t}\biggr],
\cr
}
\eqn\sixtysix
$$
where $f(t)$~is an arbitrary function satisfying~\thirtynine\ in the
first line, and we have used~\fortyfour\ in the last line. Since
$f(t)-\widetilde f(t)=O(t)$, we have set the limit of the last
integral to zero for~$\Lambda\to\infty$. This is a general formula
for arbitrary~$\widetilde f(t)$ in~\sixtytwo\
satisfiying~\thirtynine\ and reproduces the results in
Refs.~[\FRO,\AOK].\foot{%
If instead, $c_n=(-1)^n$ and $m_n=\sqrt{|n|}\Lambda$ are used
in~\sixtytwo, $\widetilde f(t)=t[\Psi((t+1)/2)-\Psi(t/2)]-1$, where
$\Psi(z)$ is the digamma function, and
$\lim_{t\to\infty}t\widetilde f(t)=1/2\ne0$. The partial integration
in~\sixtyfive\ then has an additional surface term,
$-\tr(T^aT^b)\Lambda^2g^{\mu\nu}/(32\pi^2)$ (the last integral
in~\sixtysix\ is $\log(\pi/2)$~[\AOK]). The appearance of the gauge
non-invariant piece is somewhat puzzling, because we started with a
manifestly gauge invariant expression~\fortythree. The resolution
in the present formulation seems to be the following:
When we compute the divergence of the gauge current~\fortyfive, we
encounter an integral $\int_0^\infty dt\,t\widetilde f(t)$
(see~(7.5)) which is ill-defined for this regulator function.
Therefore the WT identity~\fortyfive, and as a result, the
transverse condition of the vacuum polarization tensor, cannot be
derived with this choice.}
{}From the last expression, it is obvious that the coefficient
of~$\log\Lambda^2$, which gives the fermion contribution to the one
loop beta function, is independent of the regulator
function~\sixtytwo~[\AOK,\FUJ]. The last constant, on the other hand,
depends on the specific choice of the function (for example, for
$\widetilde f(t)=f(t)$ in~\thirtyeight, the last integral
in~\sixtysix\ is $2\log(\pi/2)-\log2$).

Let us next consider the case of anomaly-free complex
representations. In the regularized gauge current~\fiftysix, the
second term does not contribute to the vacuum polarization tensor due
to~\sixty\ ($n=2$). The same calculation as above therefore gives
\sixtysix\ with the coefficient
$\tr(\T^a\T^b)/(48\pi^2)=\tr(T^aT^b)/(24\pi^2)$, the correct result
for a chiral fermion in a complex representation~$T^a$.

As expected, the formulas \fortythree\ and~\fiftysix\ give a
transverse form without any gauge non-invariant counter terms.
Similar calculations of the fermion one loop vertex functions and a
verification of the WT identities~[\AOK] may be pursued. Since the
regularized form \fortythree\ and~\fiftysix\ for anomaly-free
representations are finite and manifestly gauge invariant, the
requirements implicit in gauge invariance should automatically be
fulfilled.
\nobreak
\chapter{Covariant anomalies}

The generalized PV regularization also provides a reliable way to
evaluate non-gauge anomalies in the anomaly-free chiral gauge
theories. Let us start with the fermion number anomaly~[\THO]: The
Majorana-type PV mass term in the formulation naturally provides the
source of the fermion number anomaly. We first take directly the
divergence of the regularized U(1) current for the real
representation~\fifty:\foot{%
If we had $f(\Dslash^2/\Lambda^2)$ instead in~\fifty\ (this regulator
is known~[\ALV] to give the consistent anomaly), we would obtain no
fermion number anomaly, $\partial_\mu\VEV{J^\mu(x)}=0$.}
$$
   \partial_\mu\VEV{J^\mu(x)}_{PV}
   =\partial_\mu\lim_{y\to x}\tr\left[
            \gamma^\mu\DslashD
            {\displaystyle1\over\displaystyle i\Dslash\DslashD}
      f(\Dslash\DslashD/\Lambda^2)\delta(x-y)\right].
\eqn\sixtyseven
$$
To evaluate this, we introduce the normalized eigenfunctions of the
hermitian operators, $\DslashD\Dslash$ and
$\Dslash\DslashD$~[\FUJIK],
$$
   \DslashD\Dslash\varphi_n(x)=\lambda_n^2\varphi_n(x),\quad
   \Dslash\DslashD\phi_n(x)=\lambda_n^2\phi_n(x),
\eqn\sixtyeight
$$
where $\lambda_n$ is real and positive. From this definition (by
appropriately choosing the phase) we see
$$
   \Dslash\varphi_n=\lambda_n\phi_n,\quad
   \DslashD\phi_n=\lambda_n\varphi_n.
\eqn\sixtynine
$$
We first use the completeness relation of $\phi_n(x)$,
$\sum_n\phi_n(x)\phi_n^\dagger(y)=\delta(x-y)$, in~\sixtyseven.
The calculation then proceeds as follows:
$$
\eqalign{
   \partial_\mu\VEV{J^\mu(x)}_{PV}
   &=\partial_\mu\left[\sum_n\phi_n^\dagger(x)
            \gamma^\mu\DslashD
            {1\over i\lambda_n^2}
      f(\lambda_n^2/\Lambda^2)\phi_n(x)\right]
\cr
   &=-i\sum_n{1\over\lambda_n}f(\lambda_n^2/\Lambda^2)
       \left[\phi_n^\dagger\rslashl\varphi_n
             +\phi_n^\dagger\rslash\varphi_n\right]
\cr
   &=-i\sum_n{1\over\lambda_n}f(\lambda_n^2/\Lambda^2)
       \left[-(\DslashD\phi_n)^\dagger\varphi_n
             +\phi_n^\dagger\Dslash\varphi_n\right]
\cr
   &=-i\sum_n
       \left[\phi_n^\dagger f(\Dslash\DslashD/\Lambda^2)\phi_n
             -\varphi_n^\dagger
              f(\DslashD\Dslash/\Lambda^2)\varphi_n
             \right]
\cr
   &=-i\lim_{y\to x}\tr
       \left[f(\Dslash\DslashD/\Lambda^2)\delta(x-y)
             -f(\DslashD\Dslash/\Lambda^2)\delta(x-y)
             \right],
}
\eqn\seventy
$$
where we have used~\sixtynine\ in several steps. The subsequent
calculation is identical to the anomaly evaluation in the path
integral framework~[\FUJIK]:
i)~$\delta(x-y)=\int d^4k\,e^{ik(x-y)}/(2\pi)^4$, ii)~shift
$e^{ikx}$ to the left, iii)~scale $k_\mu\to\Lambda k_\mu$ and expand
$f(\Dslash\DslashD/\Lambda^2)$ by~$1/\Lambda$. Finally by
using~\thirtynine, we have
$$
\eqalign{
   &\tr\int{d^4k\over(2\pi)^4}\,e^{-ikx}
   \left\{{f(\Dslash\DslashD/\Lambda^2)\atop
           f(\DslashD\Dslash/\Lambda^2)}\right\}e^{ikx}
\cr
   &\Lambdato
   {1\over4\pi^2}\int_0^\infty dt\,tf(t)\dim T\Lambda^4
   \mp{g^2\over64\pi^2}\varepsilon^{\mu\nu\rho\sigma}
   \tr\bigl(F_{\mu\nu}F_{\rho\sigma}\bigr)
   +{g^2\over48\pi^2}\tr\bigl(F_{\mu\nu}F^{\mu\nu}\bigr),
\cr
}
\eqn\seventyone
$$
and consequently,
$$
\eqalign{
   \partial_\mu\VEV{J^\mu(x)}_{PV}
   &=-i\tr\int{d^4k\over(2\pi)^4}\,e^{-ikx}
   [f(\Dslash\DslashD/\Lambda^2)
             -f(\DslashD\Dslash/\Lambda^2)]e^{ikx}
\cr
   &\Lambdato
   {ig^2\over32\pi^2}\varepsilon^{\mu\nu\rho\sigma}
   \tr\bigl(F_{\mu\nu}F_{\rho\sigma}\bigr).
\cr
}
\eqn\seventytwo
$$
This reproduces the well-known gauge invariant form of the fermion
number anomaly~[\THO].\foot{%
Since the fermion number U(1) rotation and the non-Abelian gauge
transformation commute, the Wess-Zumino consistency condition~[\WES]
implies a {\it gauge invariance\/} of the fermion number anomaly.}

We may also compute the fermion number anomaly using the WT
identity~\twentyfour. It can be verified that direct evaluation of
the right-hand side of the equation again gives the correct anomaly.
Namely,
$$
\eqalign{
   \VEV{B(x)}_{PV}&=i\lim_{y\to x}\tr\biggl(
   {\MD M\over\Dslash\DslashD+\MD M}
   -{\MprimeD\Mprime\over\Dslash\DslashD+\MprimeD\Mprime}
\cr
   &\qquad\qquad\quad
   -{\MD M\over\DslashD\Dslash+\MD M}
   +{\MprimeD\Mprime\over\DslashD\Dslash+\MprimeD\Mprime}
   \biggr)\delta(x-y).
\cr
}
\eqn\seventythree
$$
It is obvious that this leads to the last line of~\seventy, and
thus~\seventytwo.

For the anomaly-free complex representation, the regularized U(1)
current is given by~\fiftyseven. However, by repeating the above
manipulations, it is easy to see that the last, seemingly
unregularized term does not contribute to the anomaly. Namely, the
divergence of the last term identically vanishes. The divergence of
the first term of~\fiftyseven, on the other hand, gives one half
of~\seventytwo, but with the doubled gauge generator~\fiftytwo\
(note the structure constant is common for $T^a$ and $-T^{a*}$). Thus
the fermion number anomaly in the complex representation also
results in~\seventytwo, when rewritten in terms of the original
gauge generator~$T^a$.

Let us next consider the conformal anomaly represented by \fiftyone\
and~\fiftyeight. From~\seventyone, for both cases we have
$$
\eqalign{
   \VEV{T_\mu^\mu(x)}_{PV}
   &={1\over2}\tr\int{d^4k\over(2\pi)^4}\,
   e^{-ikx}[f(\Dslash\DslashD/\Lambda^2)
             +f(\DslashD\Dslash/\Lambda^2)]e^{ikx}
\cr
   &\Lambdato{1\over4\pi^2}
   \int_0^\infty dt\,tf(t)\dim T\Lambda^4
   +{g^2\over48\pi^2}\tr\bigl(F_{\mu\nu}F^{\mu\nu}\bigr),
\cr
}
\eqn\seventyfour
$$
which reproduces the correct gauge invariant result~[\CRE,\FUJIKA].
\nobreak
\chapter{Relation to the vector-like formalism}

Finally we briefly comment on the relation of the present
Lagrangian~\one, which may be called the Weyl formulation, to the
generalized PV regularization proposed by Narayanan and
Neuberger~[\NAR], the so-called vector-like formulation.

Let us begin with the complex representations. We introduce new
variables~[\AOK]
$$
   \chi_R\equiv{1+\sigma^3\over2}P_R
   \pmatrix{\psi_0\cr\psi_2\cr\vdots\cr},\quad
   \chi_L\equiv i\sigma^2{1-\sigma^3\over2}P_L\CD
   \pmatrix{\psibarT_2\cr\psibarT_4\cr\vdots\cr},
\eqn\seventyfive
$$
and
$$
   \varphi_R\equiv{1+\sigma^3\over2}P_R
   \pmatrix{\phi_1\cr\phi_3\cr\vdots\cr},\quad
   \varphi_L\equiv i\sigma^2{1-\sigma^3\over2}P_L\CD
   \pmatrix{\psibarT_1\cr\psibarT_3\cr\vdots\cr},
\eqn\seventyfivea
$$
where we have assigned even generation indices for the fermions,
and odd number indices for bosons. Then the PV Lagrangian~\one\
with \fiftythree\ and~\fiftyfive, discarding the left handed
spectator fields, is rewritten in a vector-like form of Ref.~[\NAR]
$$
   {\cal L}
   =\overline\chi\Dslash\chi
    -\overline\chi(NP_R+N^\dagger P_L)\chi
    +\overline\varphi\Dslash\varphi
    -\overline\varphi N'\varphi,
\eqn\seventysix
$$
where the covariant derivative is vector-like:
$$
   \Dslash\equiv\gamma^\mu(\partial_\mu-igA_\mu^aT^a).
\eqn\seventyseven
$$
Reflecting the fact that the original theory is chiral, the mass
matrix~$N$ has a non-trivial analytic index~[\NAR],
$\dim\ker N^\dagger N-\dim\ker NN^\dagger=1$:
$$
   N\equiv\pmatrix{
          0&2& & \cr
           &0&4 & \cr
           & &0&6\cr
           & & &\ddots&\ddots\cr
   }\Lambda,\quad
   N'\equiv\pmatrix{
   	 1& & & \cr
   	  &3& & \cr
   	  & &5& \cr
   	  & & &\ddots\cr
   }\Lambda.
\eqn\seventyeight
$$
Therefore, we see that the generalization~[\CHA] of the Weyl
formulation to arbitrary complex representations is basically
equivalent to the vector-like formulation.

How does the anomaly-free requirement emerge in the vector-like
formulation? According to Fujikawa~[\FUJ], the regularized gauge
current for~\seventysix\ is expressed as
$$
   \VEV{J^{\mu a}(x)}
   ={1\over2}\lim_{y\to x}\tr\left[
            \gamma^\mu T^a
            {\displaystyle1\over i\Dslash}
      f(\Dslash^2/\Lambda^2)\delta(x-y)\right]
    +{1\over2}\lim_{y\to x}\tr\left[
            \gamma^\mu\gamma_5 T^a
            {\displaystyle1\over i\Dslash}\delta(x-y)\right].
\eqn\seventynine
$$
Similarly, the axial U(1) current associated with
$\chi(x)\to e^{i\alpha\gamma_5}\chi(x)$ and
$\varphi(x)\to e^{i\alpha\gamma_5}\varphi(x)$ is
$$
   \VEV{J_5^\mu(x)}
   ={1\over2}\lim_{y\to x}\tr\left[
            \gamma^\mu\gamma_5
            {\displaystyle1\over i\Dslash}
      f(\Dslash^2/\Lambda^2)\delta(x-y)\right]
    +{1\over2}\lim_{y\to x}\tr\left[
            \gamma^\mu
            {\displaystyle1\over i\Dslash}\delta(x-y)\right],
\eqn\eighty
$$
and the vector U(1) current associated with
$\chi(x)\to e^{i\alpha}\chi(x)$ and
$\varphi(x)\to e^{i\alpha}\varphi(x)$ is,
$$
   \VEV{\widetilde J^\mu(x)}
   ={1\over2}\lim_{y\to x}\tr\left[
            \gamma^\mu
            {\displaystyle1\over i\Dslash}
      f(\Dslash^2/\Lambda^2)\delta(x-y)\right]
    +{1\over2}\lim_{y\to x}\tr\left[
            \gamma^\mu\gamma_5
            {\displaystyle1\over i\Dslash}\delta(x-y)\right].
\eqn\eightyone
$$

For non-anomalous gauge representations, it can be argued that the
first two currents \seventynine\ and~\eighty\ are in fact
regularized~[\NAR,\AOK,\FUJ]. Demonstration of this point requires a
somewhat detailed form of the Feynman integral and goes as follows:
The last terms in \seventynine--\eightyone\ are evaluated as
$$
\eqalign{
   &\lim_{y\to x}\tr\left[
   \gamma^\mu\left\{
   \matrix{T^a\gamma_5\cr1\cr\gamma_5\cr}
   \right\}{\displaystyle1\over i\Dslash}\delta(x-y)
   \right]
\cr
   &=-\sum_{n=0}^\infty g^n
   \int{d^4k_1\over(2\pi)^4}\cdots{d^4k_n\over(2\pi)^4}
   \int dx_1 A_{\mu_1}^{a_1}(x_1)e^{ik_1(x-x_1)}\cdots
   \int dx_n A_{\mu_n}^{a_n}(x_n)e^{ik_n(x-x_n)}
\cr
   &\quad\times\tr\left(\left\{
   \matrix{T^a\cr1\cr1\cr}
   \right\}
   T^{a_1}\cdots T^{a_n}\right)
\cr
   &\quad\times\int{d^4k\over(2\pi)^4}\tr\left(\gamma^\mu
   \left\{
   \matrix{\gamma_5\cr1\cr\gamma_5\cr}
   \right\}
   {1\over\kslash+\kslash_1+\cdots+\kslash_n}\gamma^{\mu_1}
   {1\over\kslash+\kslash_2+\cdots+\kslash_n}\gamma^{\mu_2}
   \cdots\gamma^{\mu_n}{1\over\kslash}
   \right).
\cr
}
\eqn\eightyone
$$
We then change the integration variable~$k_\mu$ to
$$
   k_\mu\to-k_\mu-k_{1\mu}-k_{2\mu}-\cdots-k_{n\mu}
\eqn\eightytwo
$$
and insert $C_D^{-1}C_D=1$ in the trace. Shifting~$C_D$ to the
right-hand side transposes the gamma matrices, and it can be
expressed as a transpose of the product of gamma matrices. Finally,
by renaming all the subscripts $(1,2,\cdots,n)\to(n,\cdots,2,1)$, we
see the integrand in~\eightyone\ is proportional to
$$
\eqalign{
   &\tr\left(\left\{
   \matrix{T^a\cr1\cr1\cr}
   \right\}
   T^{a_1}\cdots T^{a_n}\right)
   +(-1)^n\tr\left(\left\{
   \matrix{T^a\cr-1\cr1\cr}
   \right\}
   T^{a_n}\cdots T^{a_1}\right)
\cr
   &\quad\times\int{d^4k\over(2\pi)^4}\tr\left(\gamma^\mu
   \left\{
   \matrix{\gamma_5\cr1\cr\gamma_5\cr}
   \right\}
   {1\over\kslash+\kslash_1+\cdots+\kslash_n}\gamma^{\mu_1}
   {1\over\kslash+\kslash_2+\cdots+\kslash_n}\gamma^{\mu_2}
   \cdots\gamma^{\mu_n}{1\over\kslash}
   \right).
\cr
}
\eqn\eightythree
$$
For~$n\leq3$ the momentum integration is power counting divergent,
but in the {\it first two\/} lines, the coefficient for~$n\leq3$
vanishes provided that~\sixty, which is equivalent to
$\tr T^a=\tr(T^a\{T^b,T^c\})=0$, holds. Therefore for anomaly-free
representations, it can be argued that the last terms in
\seventynine\ and~\eighty\ are finite. On the other hand, the
coefficient of the last line of~\eightythree\ is proportional to
$\tr(T^aT^b)\neq0$ for $n=2$ and~$n=3$, and the above argument cannot
be applied ($J^\mu_5$ and~$\widetilde J^\mu$ are {\it different\/}
objects).

What corresponds to the unregularized U(1) current
$\widetilde J^\mu(x)$ in the Weyl formulation~\one? It is the Noether
current associated with a U(1) rotation,
$\psi(x)\to e^{i\alpha\sigma^3}\psi(x)$ and
$\phi(x)\to e^{i\alpha\sigma^3}\phi(x)$. Since the total Lagrangian
{\it is\/} invariant under this rotation, the current is conserved
($\partial_\mu\widetilde J^\mu(x)=0$) and thus cannot be used as the
fermion number current (which should be anomalous). In fact in the
Weyl formulation,
$$
\eqalign{
   &\VEV{\widetilde J^\mu(x)}_{PV}\equiv
   \VEV{\psibar\gamma^\mu\sigma^3\psi(x)
               +\phibar\gamma^\mu\sigma^3\phi(x)}
\cr
   &={1\over2}\lim_{y\to x}\tr\left[
            \gamma^\mu\sigma^3
            {\displaystyle1\over\displaystyle i\Dslash}
      f(\Dslash\DslashD/\Lambda^2)\delta(x-y)\right]
      +{1\over2}\lim_{y\to x}\tr\left[
            \gamma^\mu
            {\displaystyle1\over\displaystyle i\Dslash}
      \delta(x-y)\right],
\cr
}
\eqn\eightyfour
$$
and the last term is not regularized (note $\Dslash$ in this
expression contains $P_R=(1+\gamma_5)/2$).

We have observed that the Weyl formulation in~\S5 (complex
representation) is basically equivalent to the vector-like
formulation in Ref.~[\NAR]. An advantage of the Weyl formulation is,
however, that the requirement of the anomaly-free nature emerges in a
rather simple way; it requires only a power-counting, and the above
argument based on the change of the momentum integration variable and
the charge conjugation invariance is effectively shortcut by the
introduction of~$\T^a$, the doubled representation, and the
matrix~$\sigma^3$.

Let us now turn to real representations. It is obvious that for
real-positive representations, the Weyl formulation can be
non-equivalent to the vector-like formulation because it only
requires a finite number of regulator fields (\seventysix\ on the
other hand always requires an infinite number of such fields to have
a non-trivial analytic index). Also, for the real representations we
can construct the Lagrangian such that the seemingly unregularized
terms do not appear from the beginning (\fortythree\ and \fifty). In
the Weyl formulation, the fact that all the real representations have
no gauge anomaly can be incorporated into the Lagrangian
construction. This is another advantage of the Weyl formulation.

\chapter{Conclusion}

In this paper, we have studied the general structure of the
generalized PV regularization proposed by Frolov and Slavnov on the
basis of a regularization of composite operators. We have observed
that the PV Lagrangian provides a gauge-invariant regularization of
the chiral fermion in arbitrary anomaly-free gauge representations.
The generalization~[\CHA] to the arbitrary complex representation is
basically equivalent to the vector-like formulation in Ref.~[\NAR],
and real representations can be treated in a straightforward manner.
As the gauge current is regularized in a gauge invariant way, the
vacuum polarization tensor, for example, is found to be transverse.
We have also computed the fermion number anomaly and the conformal
anomaly within our formulation and the gauge invariant form of the
anomalies were reproduced.

A practical calculation of multi-point vertex functions is simpler if
one starts directly from the covariant regularization~[\FUJI],
because one can then choose a convenient form of the regulator
function~$f(t)$. Nevertheless, the very existence of a Lagrangian
level gauge invariant regularization makes the renormalizability and
the unitarity proofs of the anomaly-free chiral gauge theories (at
least conceptually) simpler: No gauge non-invariant counter term is
needed to compensate for the breaking of gauge symmetry by the
regularization. In particular, the standard model may be directly
treated in the scheme.

It seems to us, however, that the real importance of a possibility to
construct such a Lagrangian level gauge invariant regularization lies
in a possible implication on the lattice chiral gauge theory, in
which a consistent treatment of chiral fermions has been a long
standing problem. In fact, several proposal have been made on the
basis of the generalized PV Lagrangian~[\SLA]. A remark on this
problem is found in Ref.~[\NEW].

It would be interesting to consider a supersymmetric extension of the
generalized PV regularization, as a supersymmetric, gauge invariant
one-loop regularization.

\ack
We thank Prof.~T.~Fujiwara and H.~Igarashi for enlightening and
helpful discussions. We are grateful to Prof.~K.~Fujikawa for his
series of lectures, ``Application of path integral in quantum field
theory,'' given at Ibaraki University, which motivated the present
work. We are also grateful to Prof.~A.~A. Slavnov for clarifying some
crucial points we had misunderstood in the previous version. The work
of H.~S. was supported in part by the Ministry of Education
Grant-in-Aid Scientific Research Nos.~07740199 and 07304029.

\refout
\bye